\pdfoutput=1

\documentclass[12pt]{article}
\usepackage{jcappub}
\usepackage{pifont}
\usepackage{enumerate}
\definecolor{nicered}{rgb}{0.7,0.1,0.1}
\definecolor{nicegreen}{rgb}{0.1,0.5,0.1}
\hypersetup{colorlinks,citecolor= nicegreen,linkcolor= nicered}
\newcommand{\eqn}[1]{eq.~(\ref{#1})}
\newcommand{\eqns}[2]{eqs.~(\ref{#1})-(\ref{#2})}

\newcommand{\Eqn}[1]{Eq.~(\ref{#1})}

\newcommand{\gsim}
{\mbox{${~\raise.25em\hbox{$>$}\kern-.70em
\lower.25em\hbox{$\sim$}~}$}}
\newcommand{\lsim}
{\mbox{${~\raise.25em\hbox{$<$}\kern-.70em
\lower.25em\hbox{$\sim$}~}$}}
\newcommand{\beq}{\begin{equation}}
\newcommand{\beqa}{\begin{eqnarray}}
\newcommand{\eeq}{\end{equation}}
\newcommand{\eeqa}{\end{eqnarray}}
  \newcommand{\AddrLiege}{{IFPA, Dep. AGO, 
      Universite de Liege, Bat B5, 
Sart      Tilman B-4000 Liege 1, Belgium}}
  \newcommand{\AddrLNF}{ { INFN, Laboratori Nazionali di Frascati,
Via Enrico Fermi 40, I-00044 Frascati, Italy}}

\title{Cloistered Baryogenesis}

\author[a]{D. Aristizabal Sierra}
\author[b]{Chee Sheng Fong}
\author[b]{Enrico Nardi}
\author[b]{Eduardo Peinado}

\affiliation[a]{\AddrLiege}
\affiliation[b]{\AddrLNF}

\emailAdd{daristizabal@ulg.ac.be}
\emailAdd{Chee.Sheng.Fong@lnf.infn.it}
\emailAdd{Enrico.Nardi@lnf.infn.it}
\emailAdd{epeinado@lnf.infn.it}

\abstract{
   The cosmological matter-antimatter asymmetry can
  arise from the baryon number conserving CP asymmetry in two body
  decays of heavy particles, when the two final states carry equal and
  opposite baryon number, and one couples directly or indirectly to
  electroweak sphalerons so that its baryon asymmetry gets partly
  reprocessed into a lepton asymmetry, while the other remains
  chemically decoupled from the thermal bath with its baryon content
  frozen. After sphaleron switchoff the decay of the decoupled
  particles inject in the thermal plasma an unbalanced baryon
  asymmetry, giving rise to baryogenesis.  We highlight the features
  of this mechanism in a type-I seesaw model extended by
  adding a new colored scalar coupled to the heavy Majorana
  neutrinos.  If the colored scalar  has an ${\cal O}($TeV) mass,  it would
  leave at the LHC a characteristic signature  throughout all
  layers of the detectors.
}

\begin{document}
\maketitle
\flushbottom

\section{Introduction}
\label{sec:introduction}

The Cosmic baryon asymmetry \cite{Hinshaw:2012aka,Ade:2013zuv}
represents an indisputable evidence for physics beyond the standard
model (SM), and suggests that in the early Universe new physical
degrees of freedom must have been at work.

In the SM baryon ($B$) and lepton ($L$) number are violated only by
the $B-L$ conserving electroweak (EW) sphalerons.  In the early
Universe the rates for these processes attain thermal equilibrium at
$T\sim 10^{12}\,$GeV, and remain in equilibrium until the EW phase
transition at around $T\sim 10^{2}\,$GeV.  Any $B-L$ asymmetry generated
for example by out-of-equilibrium, $B-L$ and $CP$ violating
interactions~\cite{Sakharov:1967dj}, and surviving within this
temperature range, will then unavoidably result in a net $B$
asymmetry. This mechanism is at the basis of the standard type-I
seesaw~\cite{seesaw}  leptogenesis~\cite{Fukugita:1986hr} as well as of its 
variants~\cite{Davidson:2008bu}.  Among these variants
two realizations are particularly intriguing. In the first one, the
so-called purely flavored leptogenesis (PFL)
\cite{Nardi:2006fx,AristizabalSierra:2007ur}, the total CP asymmetry
in lepton number in the decays of the heavy Majorana neutrinos
vanishes exactly.  Leptogenesis can still proceed thanks to
non-vanishing CP asymmetries in the single lepton flavors, and thanks
to the fact that washouts violate total lepton number acting
differently along the different flavor
directions~\cite{Nardi:2006fx,FlavorEffects}.  A non-vanishing total
$B-L$ asymmetry can then result, provided all lepton-flavor-equilibrating
processes remain out of equilibrium~\cite{AristizabalSierra:2009mq}.
Another interesting variant is the so-called Dirac leptogenesis
scenario, which can yield successful baryogenesis through leptogenesis
even if $L$ remains perturbatively
conserved~\cite{Diraclepto,GonzalezGarcia:2009qd}.  In Dirac
leptogenesis~\cite{Diraclepto} heavy particle decays generate two
equal in size and opposite in sign $L$ asymmetries in the left-handed
(LH) lepton doublets and in light right-handed (RH) neutrinos. While
lepton doublets participate in EW sphaleron reactions, the RH neutrino
singlets do not. The $L$ asymmetry stored in the LH leptons is then
partially converted into a $B$ asymmetry through sphalerons
interactions.  In contrast, as long as the RH neutrinos remain
decoupled from the thermal bath, the corresponding $L$ asymmetry
remains unchanged. If decoupling holds until temperatures below EW
sphaleron freezout then, although globally $B-L=0$, a non-vanishing
$B$ asymmetry results.

Baryogenesis could also proceed via the out-of-equilibrium, C, CP and
$B$ violating decays of heavy particles, provided $L$ is conserved in
order to guarantee proton stability (see e.g. \cite{Babu:2006wz}).  It
is interesting to see if such a scenario also admits variants similar
to the ones mentioned above for leptogenesis,
and in particular  to verify if a sufficiently low scale, accessible to direct tests, 
can be reached. In this paper we show that cloistered baryogenesis does represent a viable 
scenario, although it can only work at a temperature scale above $\sim 10^7\,$GeV, 
thus  remaining out of the reach of direct tests. 

 A baryogenesis scenario similar to PFL, that is a scenario in which the total $B$-violating CP
asymmetry $\epsilon_B$ vanishes, is in general not viable.  This is because baryon
flavors, which get fully distinguished by their respective Yukawa
interactions when the temperature drops below $T \sim 10^{11}\,$GeV,
quickly equilibrate because of intergeneration mixings, driving all
baryon flavor asymmetries to zero.  Strictly speaking there is in
fact a narrow temperature window $10^{13}\;{\rm GeV} \gtrsim T \gtrsim
10^{11}\;{\rm GeV}$ when only the Yukawa reactions for the third
generation quarks are in thermal equilibrium. A third generation
baryon flavor $B_3$ then does not necessarily equilibrate with the
orthogonal flavor combination $B_{3\perp}$, so that in this window a
purely flavored baryogenesis scenario is conceivable.  However, this
appears to us as a bit cumbersome, and we will not consider further
this possibility.

The analogous of Dirac leptogenesis is instead a rather interesting
possibility.  That is, we conceive a baryogenesis scenario in which
the decays of a heavy particle do not violate either $L$ or $B$, but
an asymmetry is still generated directly in baryon number.
Essentially, two body $B$ conserving decays generate equal in size and
opposite in sign $B$ asymmetries in two different sectors. The first
one (the ``active'' sector) is coupled -- directly or indirectly -- to
EW sphalerons.  The second one remains (chemically) decoupled from the
thermal bath at least until EW sphalerons switchoff, and we will
refer to it as the uncommunicated or ``cloistered'' sector.  The
initial $B$ asymmetry stored in the active sector gets partially
converted into a $L$ asymmetry, so that its net value changes, while
the $B$ asymmetry stored in the cloistered sector remains unaffected.
After the EW phase transition heavy particles of the cloistered sector
decay, injecting their (unbalanced) baryon asymmetry in the thermal
bath, giving rise to baryogenesis.  Because of the crucial role played
by the uncommunicated  sector, we will refer to this scenario as
{\it cloistered baryogenesis}. This scenario is in fact similar in many
aspects to the so-called WIMPy baryogenesis
scenario~\cite{Cui:2011ab,Bernal:2012gv,Bernal:2013bga} in which,
however, the baryon asymmetry is generated from dark matter
annihilation instead than from heavy particle decays.

Table~\ref{tab:lepton-baryon-sector}
resumes, for the sake of illustration, the leptogenesis mechanisms
that we have briefly discussed and the corresponding baryogenesis
variants.

\begin{table}
  \centering
  \renewcommand{\arraystretch}{1.5}
  \begin{tabular}{|l|l|c||l|l|c|}
    \hline
    \multicolumn{3}{|c||}{\bf Lepton sector} & 
    \multicolumn{3}{|c|}{\bf Baryon sector} \\ \hline
    $\Delta L\neq 0$                        &
    Leptogenesis                            &
    \ding{52}                               &
    $\Delta B\neq 0$                        & 
    Direct baryogenesis                     & 
    \ding{52}                               \\ \hline
    $\epsilon_L=0$                          &
    PFL                                     &
    \ding{52}                               &
    $\epsilon_B=0$                          &
    $T \lesssim 10^{11}$ GeV          	    &
    \ding{56}                               \\ \hline
    $\Delta L=0$                            &
    Dirac leptogenesis                      &
    \ding{52}                               &
    $\Delta B=0$                            &
    Cloistered baryogenesis		            &
    \ding{52}                               \\ \hline
  \end{tabular}
  \caption{\it 
    Different mechanisms  for baryogenesis. 
    The left-hand side  lists   mechanisms 
    in which the matter-antimatter 
    asymmetry  is seeded  first in the lepton sector, and 
    $B$ is perturbatively conserved.  
    The right-hand side lists the equivalent mechanisms 
    in which the asymmetry  is seeded first   in the baryon  sector.
    In the second row $\epsilon_L=0$ and $\epsilon_B=0$ 
    refer respectively to vanishing total $L$ and $B$ violating 
    CP asymmetries. The first two mechanisms in the  baryon sector 
    require perturbative  $L$ conservation to ensure proton stability. 
    This is not required for  cloistered baryogenesis in the third row. 
    The check-mark indicate the viability of the
    corresponding scenario.
  }
  \label{tab:lepton-baryon-sector}
\end{table}
In this paper we show that cloistered baryogenesis represents a viable
scenario. We implement this mechanism in a simple extension of the
type-I seesaw that was recently put forth in
ref.~\cite{Fong:2013gaa}. In this setup, the heavy RH neutrinos $N$
couple to the $SU(2)$ singlets up-type quarks $u$, and to a new
colored scalar $\tilde u$ which, given that $N$ is a gauge singlet,
carries the same gauge quantum numbers than $u$.  In general this
scenario is not phenomenologically tenable because both $B$ and $L$
are violated and the nucleon is unstable.  However, this can be solved
by imposing exact baryon number conservation.

The rest of the paper is organized as follows. In section
\ref{sec:pert-baryogenesis-scale} we derive a lower bound on the scale
of cloistered baryogenesis. In section \ref{sec:general} we describe
the model and we discuss its phenomenological consistency. In
section~\ref{sec:b-conserving-baryogenesis} we discuss baryogenesis
within this setup, and derive the chemical equilibrium conditions and
the Boltzmann equations for baryogenesis.  In
section~\ref{sec:hypercharge} we highlight the role played in
cloistered baryogenesis by hypercharge, and finally in
section~\ref{sec:coll-phen} we present our conclusions.

\section{The temperature scale for cloistered baryogenesis}
\label{sec:pert-baryogenesis-scale}
In this section we show that assuming a non degenerate RH neutrino
spectrum, we can derive a lower bound on the temperature that allows
for successful cloistered baryogenesis.  This bound follows from the
interconnections between the CP asymmetry and the requirement that the
cloistered sector will remain uncommunicated with the active sector. While
we will be interested in the case in which a Majorana RH neutrino
decays in a SM $u$-type quark and in the complex conjugate of a new
scalar $\tilde u$ of equal baryon charge, the argument can be
presented in a more general form.

Let us consider a generic $U(1)_B$ invariant interaction between two
self conjugate particles $X_i = X_i^c$ ($i=1,2$) and other two fields
$Y$ and $Z$ carrying opposite $U(1)_B$ charges
\begin{equation}
  \label{eq:generic-Lag}
  {\cal L}=\sum_{i=1,2}\,g_i\,X_i\,Y\,Z\ + \mbox{H.c.}\ ,
\end{equation}
with $g_1$ and $g_2$ two relatively complex parameters ${\rm
  Arg(g_1^*g_2)}\neq 0$.  In general, the $X_i$ can be Majorana
fermions, with $Y$ and $Z$ a pair of complex scalar and fermion (as in
standard leptogenesis) or alternatively $X_i$ could be real scalars
and $Y,\,Z$ a pair of fermions or a pair of complex scalars (as 
in soft leptogenesis~\cite{softlepto,Fong:2011yx}).  In the
first two cases $g_i$ are dimensionless couplings, while in the last
case they have mass dimension one.  Let us now assume the mass
ordering $M_{X_2}>M_{X_1}>M_{Y}+M_{Z}$ so that the decays $X_1 \to Y\,Z,\
\bar Y\,\bar Z$ can occur. In general this decay is CP violating,
which implies a nonvanishing CP asymmetry in the number of $Y$ and $Z$
particles and antiparticles. This means that $B$ asymmetries in the
particle species $Y$ and $Z$ that are equal in size and opposite in
sign are generated.

Let us now assume that $Y$ has in-equilibrium chemical reactions with
other particles in the thermal bath, while $Z$ does not, and let us
further assume that $X_1$ decays occur before EW sphaleron freezout.
The $B$ asymmetry carried by the $Y$'s ($\Delta B_Y$) gets distributed 
between all SM particles, and because of the partial
conversion in a $L$ asymmetry through sphaleron interactions, its overall 
value is changed $\Delta B_{SM}\neq \Delta B_Y$.
In contrast, the $B$ asymmetry carried by the $Z$'s ($\Delta B_Z$)
will not change, so that eventually a net total asymmetry given by
$\Delta B=\Delta B_{SM}+\Delta B_Z$ arises.  After EW sphalerons
freezout the $Z$'s decay into SM particles (via $B$ conserving decay
modes) and this gives rise to  baryogenesis.

The CP asymmetry between the number, say, of $Y$  baryons and
$\bar Y$ anti-baryons produced in $X_1$ decays is defined as 
\begin{equation}
  \label{eq:CP-asymm-generic}
  \epsilon_{X_1}= 
  \frac{\gamma\left(X_1\to Y Z\right)
    -\gamma\left(X_1\to \bar Y\bar Z\right)}
  {\gamma^{\rm tot}
}\ ,
\end{equation}
where the $\gamma$'s are thermally averaged decay rates ($\gamma^{\rm
  tot}$ is the thermally averaged total decay width).
$\epsilon_{X_1}$ can be computed from the interference between
tree-level and one-loop vertex and wave-function diagrams.  For the
decays of Majorana fermions into a fermion/scalar pair we
have, assuming $M_{X_1} \ll M_{X_2}$~\cite{Covi:1996wh}:
\begin{equation}
  \label{eq:CP-asymm-generic-specific}
  \epsilon_{X_1}^{(fs)}   
  \simeq -\frac{\left|g_{2}\right|^{2}}{8\pi}\,
  \frac{M_{X_1}}{M_{X_2}}\sin\phi\ ,
\end{equation}
with $\phi={\rm Arg}\left[(g_1^*\,g_2)^2\right]$.  For the two 
other cases  of  scalar $X$ particles decaying into 
fermion pairs or scalar pairs, we have respectively:
\begin{eqnarray}
  \label{eq:CP-asymm-generic-scalars}
  \epsilon_{X_1}^{(ff')} &  
  \simeq& -\frac{\left|g_{2}\right|^{2}}{8\pi}\,
  \frac{M_{X_1}^2}{M_{X_2}^2}\sin\phi\ , \\
  \epsilon_{X_1}^{(ss')} &  
  \simeq& -\frac{1}{8\pi}\,
 \frac{\left|g_{2}\right|^{2}}{M_{X_2}^2}\sin\phi\ .
\end{eqnarray}
In order to maximize the CP asymmetries we  set $\sin\phi \sim 1$.
We see that in all three cases the asymmetries increase with the value
of  $g_2$.  This coupling, however, cannot become arbitrarily large
because $X_2$ mediated $YZ\leftrightarrow \bar Y\bar Z$ scatterings
would enforce  equilibrium for the $Y$ and $Z$ chemical potentials 
$\mu_Y+\mu_Z=0$ rendering cloistered baryogenesis ineffective. 
For the three cases at hand, the $2\leftrightarrow 2$ scattering rates
read:
\begin{eqnarray}
  \label{eq:two-to-two-rate-decay}
  \gamma^{(fs)}(YZ\leftrightarrow \bar Y\bar Z)
  &\simeq& \frac{1}{\pi^3}\frac{T^3}{M_{X_2}^2}
  \left|g_2\right|^4
  \to  
  \frac{64}{\pi}M_{X_1}
  \left(\epsilon^{(fs)}_{X_1}\right)^2\ , \\
  \gamma^{(ff')}(YZ\leftrightarrow \bar Y\bar Z)
  &\simeq& \frac{1}{\pi^3}\frac{T^5}{M_{X_2}^4}
  \left|g_2\right|^4  \to   \frac{64}{\pi}M_{X_1}
  \left(\epsilon^{(ff')}_{X_1}\right)^2  \ , \\
  \gamma^{(ss')}(YZ\leftrightarrow \bar Y\bar Z)
  &\simeq& \frac{1}{\pi^3}\frac{T}{M_{X_2}^4}
  \left|g_2\right|^4  \to \frac{64}{\pi}M_{X_1}
  \left(\epsilon^{(ss')}_{X_1}\right)^2\ .
\end{eqnarray}
where the limiting expressions hold for $T\to M_{X_1}$.  We see that
in all three cases the equilibrating scattering rates are proportional
to the square of the respective (maximum) CP asymmetries. Requiring
that around $T\sim M_{X_1}$ these scatterings are out of equilibrium,
that is $\gamma(YZ\leftrightarrow \bar Y\bar Z)\lsim H( M_{X_1})$
where $H( M_{X_1}) \sim 17 M_{X_1}^2/M_\text{Planck}$ parameterizes the
Universe expansion rate, yields

\begin{equation}
  \label{eq:MX1-lower-limit}
  M_{X_1}\  \gtrsim \   10^{19}\times \epsilon_{X_1}^2 \; {\rm GeV}\,.
\end{equation}
Given that values of the CP asymmetry smaller than $\epsilon_{X_1}\sim
10^{-6}$ could hardly explain the observed baryon asymmetry,
\eqn{eq:MX1-lower-limit} implies $ M_{X_1}\ \gtrsim \ 10^7\;$GeV,
which constitutes a necessary condition for successful cloistered
baryogenesis. \footnote{Similar arguments have been used in
  \cite{Racker:2013lua} to derive numerically a bound on the mass of
  the lightest RH neutrino in the inert doublet model.}

\section{General considerations}
\label{sec:general}
In the type-I seesaw model, the SM fermion sector is extended by introducing
heavy Majorana neutrinos. We assume three of them, and we denote by
$N$ the RH components $N=N_R$ while $N^c=N^c_L$ will denote the LH
components.  Besides a Majorana bilinear (mass) term $\bar N^c N$ one
can also construct a set of new fermion bilinears by coupling the
Majorana neutrinos with the SM fermions as $\bar \chi N$, where $\chi$
denotes any left-handed SM field: $\ell$, $Q$, $e^c$, $u^c$ or $d^c$
(the SM RH fields are denoted as $\chi^c=\ell^c,\,e,\,Q^c,\,d,\,u$).
The only bilinear that can be coupled in a gauge and Lorentz invariant
way without introducing additional new fields is $\bar \ell N$ because
it can be coupled to the Higgs field $\tilde H = i\sigma_2 H^*$ giving
rise to a $SU(2)\times U(1)$ invariant.  The seesaw Lagrangian, which
contains precisely this term, reads:
\begin{align}
  \label{eq:general-Lag-1}
  -{\cal L}_\text{Seesaw}&=
  \overline{\boldsymbol{\ell}}\,\boldsymbol{\lambda}
\boldsymbol{N} \tilde H +
  \frac{1}{2}\overline{\boldsymbol{N^c}}\,\boldsymbol{ M_N}
\boldsymbol{N} +\mbox{H.c.}\ .
\end{align}
Henceforth we denote matrices and vectors in boldface, so
e.g. $\boldsymbol{N}^T=(N_1,N_2,N_3)$ while ${\boldsymbol \lambda}$
and ${\boldsymbol {M_N}}$ are $3\times 3$ matrices in flavor space and,
without loss of generality, we assume that the seesaw Lagrangian
\eqn{eq:general-Lag-1} is written in the basis in which ${\boldsymbol
M_N}$ and the charged leptons Yukawa matrix are both diagonal with
real and positive entries.

Following ref. \cite{Fong:2013gaa}, by introducing new scalar fields
$\tilde \chi$ we can construct other invariants involving $N$ and the
remaining SM fermions $Q$, $e^c$, $u^c$ or $d^c$.  Clearly, since $N$
is a gauge singlet, the gauge quantum numbers of $\tilde \chi$ must
match the quantum numbers of the corresponding gauge non-singlet
fermions. In general, once these new scalars are introduced, new
operators beyond those involving the Majorana neutrinos can be
constructed by coupling $\tilde \chi$ to SM fermions
bilinears. The
resulting new Lagrangian thus has the general
form~\cite{Fong:2013gaa}:

\begin{align}
  \label{eq:general-Lag-2}
  -{\cal L}_{\tilde \chi}&= \overline{\boldsymbol{\chi}}\,
  \boldsymbol{\eta}
\,\boldsymbol{N}\, \widetilde{\chi}
  +\sum_{\boldsymbol{\chi^c}\, \boldsymbol{\chi^{\prime}}}
  \overline{\boldsymbol{\chi^c}}\,
  \boldsymbol{y}
\,\boldsymbol{\chi^{\prime}}\,
  \widetilde{\chi} + \mbox{H.c.}\ ,
\end{align}
where ${\boldsymbol \eta}$ and ${\boldsymbol y}$ are two $3\times 3$
matrices of Yukawa couplings. In the first term it is left understood
that different types of scalars have different couplings ${\boldsymbol
  \eta}= \boldsymbol\eta_{\widetilde\chi}$, while in the second term
there are different couplings also for different SM fermion bilinears
$ {\boldsymbol y}= {\boldsymbol
  y}_{\widetilde\chi}^{\chi^c\chi^{\prime}}$.

Among the various possibilities, those involving the new scalar fields
$\tilde\ell$, $\tilde e$ or $\tilde Q$ (one at the time) allow for
consistent baryon number assignments for which the ${\boldsymbol
  \eta}$ and ${\boldsymbol y}$ couplings conserve
$U(1)_B$~\cite{Fong:2013gaa}. In contrast, the inclusion of either
$\tilde u$ or $\tilde d$ yields $B$ and $L$ breaking operators and
thus, in their general form, these possibilities yield fast nucleon
decay.  Nevertheless, as we will discuss below, it is still possible to
construct viable models by imposing global $U(1)_B$ conservation as an
additional symmetry at the Lagrangian level.

\mathversion{bold}
\subsection{Adding a  $SU(2)$ singlet up-type colored scalar}
\mathversion{normal}

\label{eq:colored-scalar-scheme}

Let us now study a scenario in which a scalar field $\tilde u$ with
the same gauge quantum numbers than the RH up-type quarks is added 
to the SM plus the seesaw. The relevant new Lagrangian terms 
are:
\begin{align}
  \label{eq:setup-Lag1}
  -{\cal L}_{\tilde u}&=\overline{\boldsymbol{u}^c}\,\boldsymbol{\eta}
\,\boldsymbol{N}\,\tilde u^* +
  \overline{\boldsymbol{d}^c}\,\boldsymbol{y}
\,\boldsymbol{d}\,\tilde u + \mbox{H.c.}\ .
\end{align}
By assigning conventionally $L=0$ to the RH neutrinos,
\eqn{eq:setup-Lag1} conserves lepton number.  
However, the two terms in (\ref{eq:setup-Lag1}) cannot be made both
$U(1)_B$ invariant by any choice of the baryon charge for $\tilde u$, since 
the first term requires $B(\tilde u)=+1/3$, while the second one
requires $B(\tilde u)=-2/3$.  The simultaneous presence of $L$ and
$B$ violating terms allows the construction of operators that lead to
nucleon (${\cal N}=n,p$) decays. For example, after EW
symmetry breaking, the mixing between the heavy sterile and light active 
neutrinos results in  the $\Delta B=\Delta L=1$  dimension six operator
\begin{equation}
  \label{eq:dimension-six-operator}
  {\cal O}_6=\sqrt{\frac{m_\nu}{M_N}}\frac{\eta\,y}
  {m_{\tilde u}^2}
  \left(\overline{d^c}\,d\right)
  \left(\overline{\nu^c}\,u\right)\ ,
\end{equation}
where $m_\nu$ denotes the light neutrino mass scale,   
which induces the decay ${\cal N}\to \pi \nu$. 
The nucleon lifetime can be estimated as: 
\begin{equation}
  \label{eq:proton-lifetime}
  \tau_{\cal N} 
\sim 
  10^{32}
  \left(\frac{10^{-17}}{\eta\,y}\right)^{2}
  \left(\frac{M_N}{10^{8}\,{\rm GeV}}\right)
  \left(\frac{m_{\tilde{u}}}{1\,{\rm TeV}}\right)^{4}
  \left(\frac{0.1\,{\rm eV}}{m_\nu}\right)\,{\rm yrs}\ ,
\end{equation}
to be compared with the current bounds $\tau_{p\to
  \pi^+\nu}>0.25\times 10^{32}\,\text{yrs}$ and $\tau_{n\to
  \pi^0\nu}>1.12\times 10^{32}\,\text{yrs}$ \cite{Beringer:1900zz}.
So, if we want to keep the $\tilde u$ mass around the TeV (to allow
for its possible direct production) we see that even pushing the RH neutrino
masses $M_N \gg 10^8$ GeV, the extremely tiny size required for the
couplings would render this scenario highly unnatural.
Nucleon stability can however be guaranteed if, by imposing global
$U(1)_B$ conservation, one of the two terms in ${\cal L}_{\tilde u}$
is eliminated.\footnote{Note that one could also ensure nucleon
  stability by imposing global $U(1)_L$ conservation while allowing
  for explicit $B$ violation.  The resulting setup can allow to 
  generate  a $B$ asymmetry through out-of-equilibrium C, CP and
  $B$ violating decays of $N$, without the assistance of EW
  sphalerons.  However, in this case one has to give up the possibility of light active Majorana neutrinos.}  In the
rest of the paper we assume $B(\tilde u)=+1/3$ and  thus we drop the
second term in \eqn{eq:setup-Lag1}.

\section{Cloistered Baryogenesis}
\label{sec:b-conserving-baryogenesis}
The presence of the new interactions in (\ref{eq:setup-Lag1}) open a
new channel for RH neutrino decays: $N_i \to u_a \tilde
u^*$.\footnote{Whenever necessary we will use Latin indices $i,j,
  \dots$ to label RH neutrino generations and $a,b,\dots$ to denote
  quark flavors, while lepton flavors will be denoted by Greek indices
  $\alpha, \beta, \dots$.}  Despite being $B$ conserving, as long as
$\tilde u$ remains (chemically) decoupled from the thermal bath this
decay can provide a mechanism for baryogenesis.  Note that once the
lightest RH neutrino mass is fixed to satisfy $M_{N_1}\sim {\cal O}
(10^7\,$GeV), standard $N_1$ leptogenesis can no longer generate a
sizable B- L asymmetry because the CP asymmetry is way too
small.\footnote{Let us recall that in the temperature regime $T\sim
  10^7\,$GeV there are no directions in flavor space that remain
  protected from $N_1$ washouts, and if $N_1$ couples sizeably to all
  flavors (i.e. $|\lambda_{\alpha1}|^2 v^2/M_{N_1} \geq 10^{-3}\,$eV) any
  pre-existing asymmetry will be erased~\cite{Engelhard:2006yg}.  In
  this case our mechanism for baryogenesis could be particularly
  relevant.  Alternative possibilities for baryogenesis with
  $M_{N_1}\sim {\cal O}(10^7\,$GeV) include scenarios based on $N_2$
  leptogenesis~\cite{Vives:2005ra}, resonant
  leptogenesis~\cite{Pilaftsis:2003gt}, and models with slightly
  broken $L$~\cite{Racker:2012vw,Blanchet:2009kk}.}

\begin{figure}
  \centering
  \includegraphics{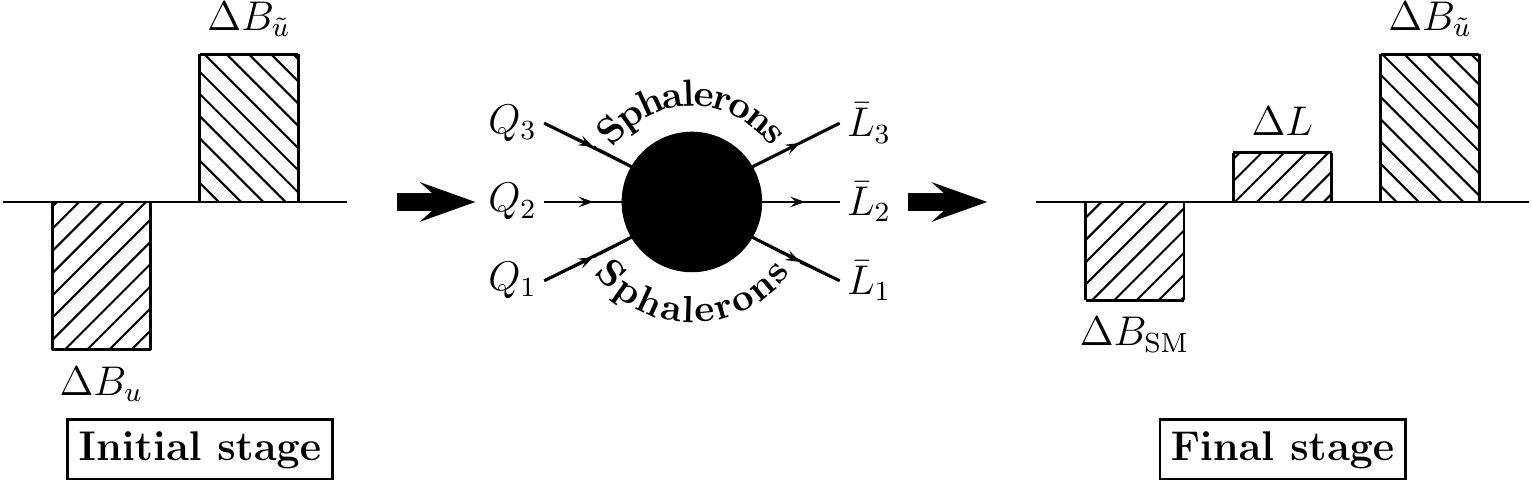}
  \caption{\it Sketch of the cloistered baryogenesis mechanism. 
    The equal
    and opposite sign $B$ asymmetries respectively in $u$ and $\tilde u$ 
    in the initial stage are denoted by $\Delta B_{u,\tilde u}$. 
At EW sphaleron decoupling
    the  $B$ asymmetry in SM particles  
     $\Delta B_{\rm SM}$ is no longer equal in magnitude 
    to the opposite sign asymmetry $\Delta B_{\tilde u}$ due to the
    EW sphaleron processes which transfer part of  the initial $\Delta B_{u}$
    to the lepton sector. A net non-vanishing  
    asymmetry $\Delta B_{\rm SM}+\Delta B_{\tilde u}\neq 0$ then results.
  }
  \label{fig:b-conserving-baryogenesis-sketch}
\end{figure}

For simplicity, let us now assume that the branching ratio for $N_1
\to \ell_\alpha H$ is much smaller than $N_1 \to u_a \tilde u^*$ so
that to a good approximation the CP asymmetry can be normalized to the
sum of the $N_1$ hadronic decays alone. In short, we assume that while
the seesaw Lagrangian still accounts for neutrino masses and mixings,
it does not have any role in baryogenesis.
Fig. \ref{fig:b-conserving-baryogenesis-sketch} illustrates how
baryogenesis can proceed in our scenario.  Initially the CP violating
out-of-equilibrium $N_1$ decays produce equal and opposite sign $B$
asymmetries in the up-type quarks ($u_a$) and in the colored scalars
($\tilde u$), that we respectively denote as $\Delta B_u$ and $\Delta
B_{\tilde u}$. In the temperature range when the decays occur ($T\sim
10^7$ GeV), EW sphaleron processes are in thermodynamic equilibrium
but, if the $\tilde u$'s remain decoupled from the hot plasma until EW
sphaleron switchoff, the asymmetry $\Delta B_{\tilde u}$ remains
unaffected. In contrast, as long as the $u$ Yukawa couplings reactions
and/or QCD sphalerons interactions are in thermal equilibrium, $\Delta
B_u$ gets first transferred to the LH quarks and eventually is
partially transformed into a $\Delta L$ asymmetry by EW sphalerons.
As a result, after EW sphaleron freeze out at $T_\text{fo}\approx
80\,\text{GeV}+0.45\,m_h\,\sim 135\,$~GeV
\cite{Burnier:2005hp,Strumia:2008cf} (for $m_h=125\,$~GeV
\cite{Aad:2012tfa,Chatrchyan:2012ufa}), a net non-vanishing $B$
asymmetry $\Delta B_\text{SM} + \Delta B_{\tilde u} \neq 0$ is
obtained, although at this stage the total $B-L$ asymmetry  still
vanishes $\Delta B_{\rm SM} + \Delta B_{\tilde u}- \Delta L =0$
(final stage in Fig.\ref{fig:b-conserving-baryogenesis-sketch}).

Now, given that astrophysical arguments rule out the possibility of
cosmologically stable heavy colored particles~\cite{Nardi:1990ku},
$\tilde u$ must eventually decay. It is a feature automatically
embedded in our model that they can do so only after EW symmetry
breaking, that is when the baryon asymmetry they release cannot be
affected any more by sphaleron interactions.  Decays occur because the
Dirac entries in the seesaw neutrino mass matrix, which are
proportional to the vev of the Higgs field, induce $N$-$\nu$ mixing,
and this opens up the decay $\tilde u \to u \nu $.  The last step of
baryogenesis thus occurs after EW symmetry breaking when $\Delta
B_{\tilde u}$ is released in the plasma. This asymmetry remains
largely unbalanced by the baryon asymmetry already present (see
section~\ref{sec:chemical-equilbrum-conditions}) and in this way a net
cosmological baryon asymmetry results.

\subsection{The viability of $B$-conserving baryogenesis}
\label{sec:viability}
\begin{figure}
  \centering
  \includegraphics{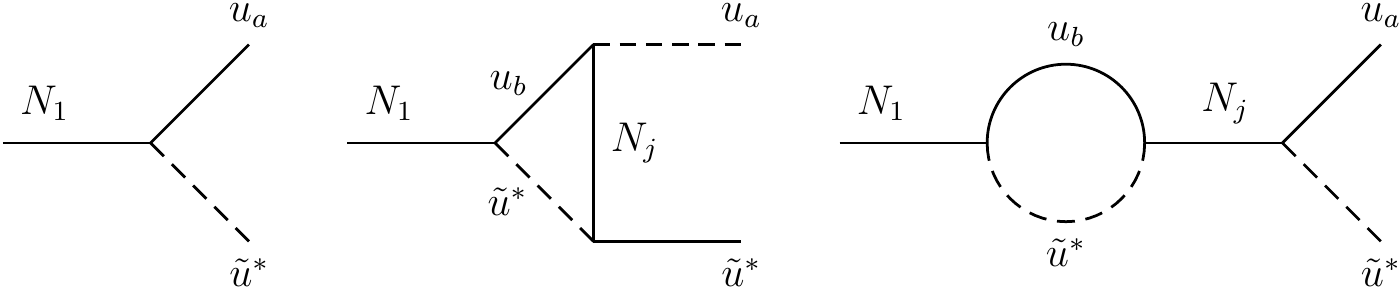}
  \caption{\it Tree-level and one-loop vertex and wave function
    corrections Feynman diagrams responsible for the CP asymmetry in
    the colored scalar scenario.}
  \label{fig:cp-asymm}
\end{figure}

The CP asymmetry in $N_1$ decays arises from the interference between
the tree-level decay and the one-loop vertex and wave function
corrections, as shown in Fig. \ref{fig:cp-asymm}. Assuming a
hierarchical RH neutrino mass spectrum ($M_{N_i}< M_{N_j}$ for $i<j$),
summing over quark flavors and taking into account color factors, the
CP asymmetry between the number of $\tilde u$ and $\tilde u^*$ scalars
produced in $N_1$ decays is
\begin{equation}
  \label{eq:CP-asymm-model-dependent}
  \epsilon^{\tilde u}_{N_1}\simeq -\frac{1}{4\pi}
  \frac{1}{(\boldsymbol{\eta}\boldsymbol{\eta}^\dagger)_{11}}
  \sum_{j\neq 1}\mathbb{I}\mbox{m}
  \left[\left(\boldsymbol{\eta}\boldsymbol{\eta}^\dagger\right)^2_{1j}\right]
  \frac{M_{N_1}}{M_{N_j}}\ .
\end{equation}
In addition to a sufficiently large CP asymmetry, 
the success of cloistered baryogenesis
requires that the following three conditions
are satisfied:
\begin{itemize} 
\item[ ($i$)] The decays $\Gamma_{N_1}\equiv \Gamma(N_1\to \sum_a
  u_a\tilde u^*) $ should occur out of equilibrium, and the rate for
  the $N_1$ semileptonic decays should satisfy $\Gamma(N_1\to
  \sum_\alpha \ell_\alpha H^*) < \Gamma_{N_1}$;
\item[($ii$)] The scalars $\tilde u$ should remain chemically
  decoupled from the thermal bath (of course strong interactions will
  keep them in kinetic equilibrium);
\item[($iii$)] The decays $\tilde u\to u\nu$, which eventually fix the
  final amount (and sign) of the baryon asymmetry, should occur well
  before the Big Bang Nucleosynthesis (BBN) era.
\end{itemize} 
All together these conditions enforce constraints on the relevant
model parameters.

Condition ($i$) is satisfied provided that  
$\Gamma_{N_1} = \frac{3}{8\pi}|\eta_{a1}|^2 M_{N_1}$
is smaller than the Universe expansion rate at $z\equiv
M_{N_1}/T\sim 1$, which implies
  \begin{equation}
    \label{eq:gamma-decay-expansion-rate}
    |\eta_{a1}| \lesssim 1\times 10^{-5}
    \left(\frac{M_{N_1}}{10^7\,\text{GeV}}\right)^{1/2}\, .
\end{equation}
Here and henceforth we normalize the RH neutrino mass to
$10^7\,$~GeV, as suggested  by the condition 
for successful baryogenesis discussed in
section~\ref{sec:pert-baryogenesis-scale}.

Condition ($ii$) implies specific requirements on the rates of the $s$
and $t$ channel scattering process $u_a \tilde u^*\leftrightarrow \bar
u_b \tilde u$, on the (inverse) decay rates of the heavier RH neutrinos
 $u_a \tilde u^*\to N_{2,3}$ , and on the rates of the three-body decays
$\tilde u\to \ell_\alpha\,H\,u_a$:
\begin{itemize}
\item {\bf $N_{2,3}$ mediated $s$ and $t$ channel
    $2\leftrightarrow 2$ scatterings}. As argued in
  section~\ref{sec:general}, $2\leftrightarrow 2$ processes can place
  tight constraints on baryogenesis.  The role played in our specific
  case by $N_{2,3}$ mediated $u_a\,\tilde u^*\leftrightarrow \bar
  u_b\,\tilde u$ scatterings can be readily understood from the
  one-loop diagrams in Fig. \ref{fig:cp-asymm}, since they involve the
  same couplings as the $2\leftrightarrow 2$ reactions.  Requiring
  that these reactions are out of equilibrium enforces constraints on
the ratio between 
the Yukawa couplings and the heavier neutrino masses, and in turn 
this can imply a too large suppression of the CP asymmetries. 
 Considering for definiteness only $N_2$ and one   single channel,  the
  $u_a\,\tilde u^*\leftrightarrow \bar u_b\,\tilde u$ scattering rate
  is approximately given by 
  \begin{equation}
    \label{eq:scttering-rate-colored-scalar-scenario}
    \Gamma(u_a\,\tilde u^*\leftrightarrow \bar
    u_b\,\tilde u)\simeq \frac{1}{\pi^3}
    \frac{M_{N_1}^3}{M_{N_2}^2}
    \left|\eta_{a2}\right|^2
    \left|\eta_{b2}\right|^2\ , 
  \end{equation}
  and demanding that this reaction to be decoupled at $z \sim 1$, implies the
  following constraint on the Yukawa couplings:
  \begin{equation}
    \label{eq:constraint-from-scatterings}
    \left|\eta_{a2}\right|\,\left|\eta_{b2}\right|\lesssim
    2\times 10^{-5}\,\left(\frac{M_{N_2}}{M_{N_1}}\right)\,
    \left(\frac{M_{N_1}}{10^7\,\text{GeV}}\right)^{1/2}\ .
  \end{equation}
The analogous limits for $N_3$ mediated reactions  are obtained  by 
substituting  $\eta_{a2} \to \eta_{a3}$  and $M_{N_2} \to M_{N_3}$.

\item {\bf $N_{2,3}$ inverse decays}: At $T\sim M_{N_1}\ll
  M_{N_{2,3}}$, $N_{2,3}$ inverse decays are Boltzmann suppressed, but one has to ensure that this
  suppression is sufficient to avoid depleting the asymmetry from
  $N_1$ decays.  The thermally averaged inverse decay rate can be
  approximately written as
  \begin{equation}
    \label{eq:inverse-decays}
    \gamma(u_a\,\tilde u^*\to N_{2,3})\simeq \Gamma(N_{2,3}\to u_a\,\tilde u^*)\,
    \left(\frac{M_{N_{2,3}}}{M_{N_1}}\right)^{3/2}\,e^{-M_{N_{2,3}}/T}\ .
  \end{equation}
In terms of the RH neutrino masses and Yukawa couplings,
  the condition $\gamma(u_a\,\tilde u^*\to N_{2,3})\lesssim\,H(z \sim
  1)$ translates into
  \begin{equation}
    \label{eq:constraint-inverse-decays}
    |\eta_{a(2,3)}|\lesssim \ 1.5\times 10^{-5}\,
    \left(\frac{M_{N_1}}{M_{N_{2,3}}}\right)^{5/4}\,
    \left(\frac{M_{N_1}}{10^7\,\text{GeV}}\right)^{1/2}\,
    e^{M_{N_{2,3}}/2M_{N_1}}\ .
  \end{equation}
  For example, by taking $M_{N_1}=10^7$ GeV and $M_{N_1}/M_{N_j} =
  0.04$, we obtain $|\eta_{aj}| \lesssim 7\times 10^{-2}$.  Because of
  the exponential factor, as soon as the ratio $M_{N_1}/M_{N_j}$
  falls below $\sim 10^{-2}$ this constraint becomes completely
  irrelevant with respect to the constraints from $2\leftrightarrow 2$
  scatterings~\eqn{eq:constraint-from-scatterings}, which are only
  power suppressed.

\item {\bf Three-body decays}: Already above the EW symmetry breaking
  scale, the colored scalars can decay via the RH-neutrino-mediated
  three body channel $\tilde u\to u_a\,\ell_\alpha H$. If sufficiently
  fast, this process would spoil the generation of the 
baryon asymmetry because $\Delta B_{\tilde u}$ is re-injected in the thermal
bath too early, that is when EW sphalerons are still active.  
This decay channel, however, involves not only the $\boldsymbol{\eta}$
  couplings but also the parameters responsible for neutrino masses
  and mixings $\boldsymbol{\lambda}$. The corresponding constraint reads
  \begin{equation}
    \label{eq:three-body-constraints}
    \left|\lambda_{\alpha j}\right|\left|\eta_{a j}\right|\lesssim
    2\times 10^{-3}
    \left(\frac{T_\text{fo}}{135\,{\rm GeV}}\right)
    \left(\frac{M_{N_j}}{10^{7}\,{\rm GeV}}\right)
    \left(\frac{1\,{\rm TeV}}{m_{\tilde{u}}}\right)^{3/2}\ .
  \end{equation}
  For ${\cal O}(M_{N_j})\sim 10^7\,$~GeV 
  consistency with a neutrino mass scale below, say, a few tenths of eV   
  already requires $|\lambda| \lesssim 10^{-3}$, so that the constraint
  \eqn{eq:three-body-constraints} is easily satisfied and basically of
  no importance.

\end{itemize}

After EW symmetry breaking the active-RH neutrino mixing induces the
decays $\tilde u\to u_a\,\nu$ which release the asymmetry $\Delta
B_{\tilde u}$ in the thermal bath.  Condition ($iii$) requires that
these decays occur at temperatures safely above the temperature where
the $n/p$ ratio freezes out and BBN starts.  Note that BBN constraints
on hadronically decaying massive particles \cite{Kawasaki:2004qu}
assume in general that no baryon asymmetry is generated in these
decays, and thus involve a different type of effects.
In our case the requirement that has to be imposed is that the correct
value (and sign) of the baryon-to-photon ratio is established as the
initial condition for BBN.  This yields the following constraint:
\begin{equation}
  \label{eq:BBN-constraint}
  \left|\eta_{aj}\right|\gtrsim
  3\times 10^{-4}\,\left(\frac{T_\text{BBN}}{10\,\text{MeV}}\right)\,
  \left(\frac{M_{N_j}}{10^7\,\text{GeV}}\right)^{1/2}\,
  \left(\frac{0.1\,\text{eV}}{m_\nu}\right)^{1/2}\,
  \left(\frac{1\,\text{TeV}}{m_{\tilde u}}\right)^{1/2}\ .
\end{equation}
Note that the out-of-equilibrium condition require the $\eta_{a1}$
couplings to be smaller than $\sim 10^{-5}$ (see
\eqn{eq:gamma-decay-expansion-rate}). As shown in
\eqn{eq:constraint-from-scatterings}, constraints on the
$\eta_{a(2,3)}$ couplings are much weaker, implying that  $\tilde u\to
u_a \nu$  decays can occur at a sufficiently early stage   thanks to the contributions
from $\nu-N_{2,3}$ mixing. For example, 
fixing the $\eta_{a(2,3)}\sim 0.03$ and $M_{N_{(2,3)}}\sim 10^8$ GeV, one obtains   
for the colored scalar  a  mean lifetime  $\tau_{\tilde u}\sim 10^{-4}\,$ sec. 
which ensures that  all $\tilde u$ will have decayed much before BBN.

\subsection{Chemical equilibrium conditions and kinetic equations}
\label{sec:chemical-equilbrum-conditions}

A more quantitative analysis of cloistered baryogenesis requires
writing down the relevant Boltzmann equations, while taking into
account the chemical equilibrium conditions enforced by those
reactions that, in the range of temperatures relevant for the
production of $\Delta B_{\rm SM}$ and $\Delta B_{\tilde u}$, are
faster than the Universe expansion rate.  In the following we fix this
temperature at $T\sim 10^7\,$~GeV that, as was discussed in
section~\ref{sec:pert-baryogenesis-scale}, within our scenario is the
lowest value compatible with successful baryogenesis.

We start by recalling some well known relations and by introducing
notations.  The number density asymmetry of bosons  and fermions
 $\Delta n_{b,f}\equiv n_{b,f}-\bar{n}_{b,f}$ is related to
the corresponding chemical potentials $\mu_{b,f}$.  In the
relativistic limit ($m_{b,f}\ll T$) and at first order in
$\mu_{b,f}/T\ll1$ the corresponding relations read:
\begin{eqnarray}
\Delta n_{b} & = & \frac{T^3}{3}\left(\frac{\mu_{b}}{T}\right),
\;\;\;\;\;\Delta n_{f}=\frac{T^3}{6}\left(\frac{\mu_{f}}{T}\right)\, .
 \label{eq:numden_chempot}
\end{eqnarray}
Note that above we have defined $\Delta n_{b,f}$ 
as particle number asymmetries for degree of freedom.   
Then the number of degrees of freedom $g_{b,f}$ of each particle 
has to be taken into account when constructing global 
asymmetries for example  in baryon or  in lepton number. 
To remove the effect of the expansion of the Universe it is customary
to normalize the particle number densities to the entropy density
$s=g_{\star}\left(2\pi^{2}/45\right)T^{3}$ i.e. $Y_{\Delta
  n}=Y_{n}-Y_{\bar{n}}\equiv \Delta n/s$.
In principle, for each non-self conjugate particle there is one
chemical potential.  However, the overall number of independent chemical 
potentials is drastically
reduced by the different constraints imposed by the chemical
equilibrium conditions and/or conservation laws, and eventually 
it turns out to be equal to the number of conserved charges.  We follow
ref. \cite{AristizabalSierra:2009mq} and adopt the notation
$X\equiv\mu_X$, where $X$ is either a SM field or $\tilde u$ and
$\mu_X$ its corresponding chemical potential. The constraints on the
chemical potentials are:

\begin{enumerate}[1)]
\item \label{chemical-condition1} 
Chemical potentials for the gauge
  bosons vanish $W=B=g=0$ and hence all the particles belonging to the
  same $SU(3)_{C}$ or $SU(2)_{L}$ multiplets have the same chemical
  potential~\cite{Harvey:1990qw}.

\item \label{chemical-condition2} 

  The Yukawa reactions for the second and third generations of SM
  fermions are in thermodynamic equilibrium.  For simplicity we assume
  equilibrium also for the first generation (numerical differences do
  not exceed the ten percent level~\cite{Nardi:2005hs}).
  Also, intergenerational quark mixing ensures $Q_a=Q$, so we get:
  \begin{eqnarray}
    \label{eq:Yukawa-chemical-Eq-condition-Lep}
    \ell_{\alpha}-e_{\alpha}-H & = & 0 \qquad (\alpha=e,\mu,\tau)\ ,\\
    \label{eq:Yukawa-chemical-Eq-condition-up}
    Q-u_{a}+H & = & 0 \qquad (u_{a}=u,c,t)\ ,\\
    \label{eq:Yukawa-chemical-Eq-condition-down}
    Q-d_{a}-H & = & 0 \qquad (d_{a}=d,s,b)\ .
\end{eqnarray}

\item \label{chemical-condition4}  Equilibrium of  EW sphaleron
  interactions yields 
  \begin{eqnarray}
    \label{eq:EW+QCD-chemical-Eq-cond}
    9Q+\sum_{\alpha}\ell_{\alpha} & = & 0.
  \end{eqnarray}

\item \label{chemical-condition3} In terms of chemical potentials,  the condition of cosmological hypercharge neutrality 
$\sum_X {\cal Y}_X g_X \Delta n_X $ (with ${\cal Y}_X$ the $X$-particle hypercharge and $g_X$ its number of degrees of freedom) translates into:
\begin{align}
  \label{eq:hypercharge-neutrality}
  3Q + \sum_a\left(2 u_a - d_a\right)
  -\sum_\alpha\left(\ell_\alpha + e_\alpha\right)
  +2 H  + 4 \tilde u =0\ .
\end{align}
\end{enumerate}
Note that when all quarks Yukawa reactions are assumed to be in
thermodynamic equilibrium QCD sphalerons do not impose an independent
constraint~\cite{Nardi:2005hs}.
All in all, the initial 15 chemical
potentials $u_a, d_a, e_{\alpha}, \ell_{\alpha}, Q, H, \tilde{u}$, are
reduced to 4 by the $9+1+1=11$ conditions implied by
\ref{chemical-condition2}, \ref{chemical-condition4}  and \ref{chemical-condition3}, namely, by Eqs. (\ref{eq:Yukawa-chemical-Eq-condition-Lep}-\ref{eq:Yukawa-chemical-Eq-condition-down}), (\ref{eq:EW+QCD-chemical-Eq-cond}) and (\ref{eq:hypercharge-neutrality}). 
 As mentioned above, this
could have been expected simply from symmetry considerations.  In the
approximation in which the $N_1 \leftrightarrow u_a \tilde u^*$
reactions are completely out of equilibrium, there are four conserved
charges corresponding to global $U(1)_{\tilde u}$~\footnote{Note that
  the presence of a global $U(1)_{\tilde u}$ can be in fact taken as
  an \emph{operative} definition of having $\tilde u$ decoupled from
  the thermal plasma.}  and to the three global $U(1)_{\Delta_\alpha}$
where $\Delta_\alpha \equiv \Delta B_{\rm SM}/3-\Delta L_\alpha$. Hence
the normalized number density asymmetries of all particle species 
can be expressed in terms of the asymmetries in the four charges 
$Y_{\Delta_{\tilde u}}$ and $Y_{\Delta_\alpha}$. We obtain:
\begin{align}
  \label{eq:remaining-chemical-potentials}
Y_{\Delta \ell_\alpha} &= -\frac{3}{79} Y_{\Delta \tilde u} 
+ \frac{16}{711} 
Y_{\Delta(B_{\rm SM} - L)}
-   \frac{1}{3} Y_{\Delta_\alpha}\,,  
\qquad 
  Y_{\Delta u_a} =
-\frac{12}{79} Y_{\Delta \tilde u} -\frac{5}{237} 
Y_{\Delta(B_{\rm SM} - L)}\,, 
 \nonumber\\
Y_{\Delta e_\alpha} &= \frac{10}{79} Y_{\Delta \tilde u} 
+ \frac{52}{711}
Y_{\Delta(B_{\rm SM} - L)}
-   \frac{1}{3} Y_{\Delta_\alpha}\,,   
\qquad   \ \ 
 Y_{\Delta d_a} =
\frac{14}{79} Y_{\Delta \tilde u} 
+\frac{19}{237} 
Y_{\Delta(B_{\rm SM} - L)}\,, 
  \nonumber\\
  Y_{\Delta Q} &=
\frac{1}{79} Y_{\Delta \tilde u} +\frac{7}{237} 
Y_{\Delta(B_{\rm SM} - L)}\,, 
\qquad \qquad \qquad  \;
 Y_{\Delta H} =
-\frac{26}{79} Y_{\Delta \tilde u} 
-\frac{8}{79} 
Y_{\Delta(B_{\rm SM} - L)}\,, 
\end{align}
where $Y_{\Delta(B_{\rm SM} - L)} = \sum_{\alpha} Y_{\Delta_\alpha}$.
It is important to note, as could be readily verified 
from the previous relations, that since the hypercharge 
condition \eqn{eq:hypercharge-neutrality} is different 
from the SM case because of the presence of the contribution 
from the $\tilde u$ scalars, the relation between the amount 
of baryon  asymmetry and $B-L$ asymmetry 
stored in SM particles is also changed, and reads 
\begin{equation}
\label{eq:DeltaBSM} 
Y_{\Delta B_{\rm SM}} = \frac{28}{79}Y_{\Delta(B_{\rm SM} - L)} + 
 \frac{12}{79} Y_{\Delta \tilde u} \,,
\end{equation}
where the first term on the RH side is the usual SM result, while   
the second  is new. 

Now, the dynamical equations for baryogenesis get largely simplified in the
approximation in which $N \leftrightarrow \ell_\alpha H$ interactions
are neglected, and we will adopt this approximation in the last part
of this section.  Although at $T\sim 10^7\,$GeV the three lepton
flavors are neatly distinguished by their Yukawa
interactions~\cite{Nardi:2006fx,FlavorEffects}, in this
approximation all dynamical processes become symmetric under a
relabeling of the lepton flavor index $\alpha$, and this means that
the asymmetries $Y_{\Delta_\alpha}$ evolve in the same way and must be
equal at all times.  Thus we can simply set $Y_{\Delta_\alpha}=
\frac{1}{3}Y_{\Delta(B_{\rm SM} - L)}$.
Another simplification stems from the fact 
that  at this stage the total $B-L$ is a conserved quantity, that is 
\begin{equation}
\label{eq:DeltaBmL} 
Y_{\Delta(B_{\rm SM} - L)} + 
Y_{\Delta \tilde u} =0\,, 
\end{equation}
where the second term in the RH side is the contribution to  total
$\Delta B$ from the $\tilde u$ scalars.  \Eqn{eq:DeltaBmL} implies
that to estimate the baryon asymmetry yield of cloistered baryogenesis
is sufficient to solve a system of just two Boltzmann equations:
\begin{align}
  \label{eq:kinetic-Eqs1}
  \dot Y_{N_1}&=-(y_{N_1}-1)\,\gamma_{N_1}\ ,
  \\
  \label{eq:kinetic-Eqs2}
  \dot Y_{\Delta \tilde u}&=
  (y_{N_1}-1)\epsilon^{\tilde u}_{N_1}\,\gamma_{N_1}
  +\frac{1}{2}
  \left(y_{\Delta u}-y_{\Delta\tilde u}\right)\,\gamma_{N_1}\ ,
\end{align}
where $\gamma_{N_1}$ denotes the thermally averaged $N_1$ decay rate,
the time derivative is defined as $\dot Y\equiv sHz\,dY/dz$, the
density asymmetries have been normalized as $y_{\Delta \tilde u}=
Y_{\Delta \tilde u}/Y^{\rm Eq}_{b}$ and $y_{\Delta u} = Y_{\Delta
  u}/Y^{\rm Eq}_{f}$ with the respective boson and fermion equilibrium
abundances $Y_{b}^{\rm Eq}=2Y_{f}^{\rm Eq}=\frac{15}{4\pi^{2}g_{*}}$, we have
dropped the RH up-type quark flavor index by setting, according
to~\eqn{eq:remaining-chemical-potentials}, $u_a = u$, and finally we
have neglected on-shell and off-shell contributions from $N_{2,3}$.
Note that $y_{\Delta u}$ appearing in the washout term
in~\eqn{eq:kinetic-Eqs2} has to be evaluated by means of the first
relation in the right side column in
\eqn{eq:remaining-chemical-potentials} together with
\eqn{eq:DeltaBmL}.  This yields 
\begin{equation}
y_{\Delta u}=-\frac{62}{237}\, y_{\Delta  \tilde u}.
\end{equation}

According to \eqn{eq:DeltaBSM} and \eqn{eq:DeltaBmL}, once the era of
$N_1$ decays is ended, but before the $\tilde u$ scalars start
decaying (let us say, for definiteness, at temperatures around the EW
phase transition), the amount of baryon asymmetry stored in SM
particles is
\begin{equation}
\label{eq:DeltaBSMfo}
Y^{\rm EW}_{\Delta B_{\rm SM}}= 
- \frac{16}{79} Y^{\rm EW}_{\Delta \tilde u}\, ,
\end{equation}
that is  about 20\% of the final value of $Y_{\Delta \tilde u}$ but  of 
opposite sign.
However, what should be confronted with cosmological measurements is
the baryon asymmetry {\it after} all the $\tilde u$ scalars have
decayed (say, for definiteness, at temperatures around the BBN era)
which is given by:
\begin{equation}
\label{eq:DeltaBSMBBN}
Y^{\rm BBN}_{\Delta B}= 
Y^{\rm EW}_{\Delta B_{\rm SM}} + 
Y^{\rm EW}_{\Delta \tilde u} =
\frac{63}{79}  Y^{\rm EW}_{\Delta \tilde u}\, .
\end{equation}
Confronting \eqn{eq:DeltaBSMfo}  and \eqn{eq:DeltaBSMBBN}  shows that
the main contribution to the present cosmological baryon asymmetry
as well as  its sign,   are determined  by the asymmetry stored in the colored 
scalars $\tilde u$, which remain decoupled from the thermal bath down to
temperatures well below the EW phase transition.  This asymmetry could in fact
be released at temperatures as low as ${\cal O}(10$ MeV),  right 
before  the onset of BBN.

\section{The role of hypercharge}
\label{sec:hypercharge}

The analysis of the previous section indicates that in our
baryogenesis model the small amount of perturbative $L$ violation does
not play any crucial role. As regards baryon number, apart from
sphaleron interactions, at the Lagrangian level it remains conserved
at all stages.  It is then interesting to ask which is the fundamental
charge whose asymmetry is feeding all particle asymmetries, and
eventually baryogenesis. As we will now argue, the answer is that this
role is played by the asymmetry in the total hypercharge of the SM
particles.\footnote{That such an asymmetry could drive baryogenesis
  was noted already long ago in ref.~\cite{Antaramian:1993nt}.}  The
following example will help to make this point more clear.  Let us
assume the following setup:

\begin{itemize}
\item The two baryon asymmetries $\Delta B_{\rm SM}$ and $\Delta
  B_{\tilde u}$ are generated in the out-of-equilibrium decays of
  $N_2$, with the usual condition $\Gamma(N_2 \to \ell H)\ll
  \Gamma(N_2 \to u \tilde u^*)$\,.

\item The $N_1$ decay rate $\Gamma(N_1 \to u \tilde u^*)$ is instead
  negligible, while the $L$  violating decays and inverse decays $N_1
  \leftrightarrow \ell H, \bar \ell H^*$ are in full thermal
  equilibrium.
\end{itemize}

This second assumption implies one additional condition, which   should be added 
to the  set of chemical potential relations  
\eqns{eq:Yukawa-chemical-Eq-condition-Lep}{eq:EW+QCD-chemical-Eq-cond}.
Recalling that the Majorana states $N$ have vanishing chemical potential, this 
condition reads: 
\begin{equation}
\label{eq:LH0}
\ell_\alpha + H = 0   \qquad (\alpha =e,\mu,\tau)\,.
\end{equation}
Now, from the hypercharge neutrality condition
\eqn{eq:hypercharge-neutrality} we have that the sum of the SM
particle number asymmetries weighted by the hypercharge of the
corresponding particles, and  written in terms of chemical potentials, 
should add to $-2 g_{\tilde u} \,{\cal Y}_{\tilde u}\,{\tilde u}= - 4
\tilde u$ (with $g_{\tilde u}=3$ the color degrees of freedom of
$\tilde u$, and ${\cal Y}_{\tilde u}=2/3$ its hypercharge), that is it
should exactly balance the amount of hypercharge asymmetry stored in
the cloistered sector.  The solution of the set of chemical potential
conditions
\eqns{eq:Yukawa-chemical-Eq-condition-Lep}{eq:EW+QCD-chemical-Eq-cond}
and \eqn{eq:LH0} is straightforward: since all SM reactions as well as
 $N_1 \leftrightarrow \ell H, \bar \ell H^*$ conserve exactly
hypercharge, the chemical potential of the SM particles must be simply
proportional to the particle hypercharges:
\begin{equation}
\label{eq:proportional}
\mu_\phi = \kappa\, {\cal Y}_\phi   
 \qquad (\phi =\ell_\alpha , e_\alpha , Q ,u_a , d_a, H)\,. 
\end{equation}
The coefficient  $\kappa$ can then be directly evaluated 
from total hypercharge conservation:
\begin{equation}
\label{eq:kappa}
\kappa  = - \frac{2\, g_{\tilde u}\,{\cal Y}_{\tilde u}}
{\sum_\phi\,  g_\phi\,{\cal Y}_{\phi}^2} \, 
\tilde u\,. 
\end{equation}
Note that, with a slight abuse of notation,  within  the sum in the 
denominator   $g_H = 2 \times 2$ where the first factor is from 
the Higgs $SU(2)$ degrees of freedom, and the second 
from boson/fermion statistics $\Delta n_H/\Delta n_f = 2 \mu_H/\mu_f$. 
This allows us to write the chemical potentials of all the SM
particles in terms of $\tilde u$, which in turn is obtained by
integrating the Boltzmann equations
(\ref{eq:kinetic-Eqs1}) and (\ref{eq:kinetic-Eqs2}).  We thus see that even when
$L$ is violated by in-equilibrium reactions and $B$ is perturbatively
conserved, still, in order to balance the net amount of hypercharge
stored in the cloistered sector, all SM particles carrying hypercharge
develop non-vanishing asymmetries.

\section{Conclusions}
\label{sec:coll-phen}

We have studied a scenario where the cosmological matter/antimatter
asymmetry stems from an asymmetry in baryon number related to heavy
particle decays.
To ensure nucleon stability, baryon number is imposed as a symmetry at the Lagrangian
level; however,  baryogenesis   can still proceed
because a certain amount of baryon asymmetry generated from the $B$
conserving decays of heavy particles is confined into a {\it
  cloistered} sector that remains chemically decoupled from the
thermal bath until $B+L$ violating sphaleron reactions are switched
off.  An initial equal amount of baryon asymmetry stored in the SM
sector gets instead partially transformed into a lepton
asymmetry. When the asymmetry in the cloistered sector is eventually
released into the thermal bath (in our model this can naturally occur
at temperatures not far above the onset of BBN) the unbalance between the two
asymmetries gives rise to baryogenesis.

We have studied some necessary conditions to allow for successful
baryogenesis within this scenario.  For example we have found that
sufficiently large CP asymmetries together with the requirement that
the cloistered sector will remain chemically decoupled from the  
thermal bath, require  that the mass of the initial heavy decaying particles 
must be at least of ${\cal O}\sim 10^7\,$~GeV.
While this is about two orders of magnitude lower than  the scale required for successful leptogenesis, it remains well above    the TeV scale, thus excluding the possibility of direct  tests at colliders.

We have implemented cloistered baryogenesis within a specific setup,
based on a straightforward extension of the standard seesaw model to
which a colored scalar $\tilde u$ with the same quantum numbers of
the up-type RH quarks is added.  We have illustrated in detail the
viability of this realization, we have analyzed various constraints
showing that they can all be satisfied, we have derived the chemical
equilibrium conditions that relate the SM particle asymmetries, and
we have written down the kinetic equations whose solution
allows to estimate the present amount of cosmological baryon
asymmetry. Finally, we have highlighted the fundamental role played 
in our setup by  hypercharge conservation~\cite{Antaramian:1993nt}. 

If the new colored states  which are the clue ingredient of  cloistered
baryogenesis have, as we have assumed, masses of ${\cal O}($TeV), they
would be well within the LHC reach even with moderate luminosity,
given that their production rates are   governed by $\alpha_s$.  The
requirement that they will keep decoupled from the thermal bath
implies, as a specific signature, a relatively long lifetime.  Thus,
they could be produced at the LHC  in large numbers, and leave a characteristic
signature throughout all layers of the detectors, much alike the long
lived colored particles studied in~\cite{Buckley:2010fj}.  The
experimental observation of colored scalars  with these characteristics
will clearly not suffice to identify cloistered baryogenesis as the mechanism 
responsible for the cosmic baryon asymmetry, but it would certainly
support this idea.

\section{Acknowledgments}
DAS wants to thank the ``Laboratori Nazionali di Frascati'' for
hospitality during the completion of this work. DAS is supported by
the Belgian FNRS agency through a ``Charg\'e de Recherche'' contract.
CFS would like to thank the hospitality of IFPA, University of Li\`{e}ge
where part of this work was carried out.

\end{document}